\documentclass[10pt,superscriptaddress,amsmath,amssymb,aps,prl,twocolumn,floatfix]{revtex4-1}

\usepackage{graphicx}% Include figure files
\usepackage{dcolumn}% Align table columns on decimal point
\usepackage{float}
\usepackage{bm}% bold math
\usepackage[dvipsnames]{xcolor}

\begin{document}

\preprint{APS/123-QED}

\title{Nonlinear Photon Pair Generation in a Highly Dispersive Medium}

\author{David J.~Starling}
\affiliation{Division of Science, Pennsylvania State University, Hazleton, PA 18202, USA}
\author{Jacob Poirier}
\affiliation{School of Physics and Astronomy, Rochester Institute of Technology, Rochester, NY 14623, USA} \author{Michael Fanto}
\affiliation{Air Force Research Laboratory, Rome, New York 13441, USA}
\affiliation{Microsystems Engineering, Rochester Institute of Technology, Rochester, New York 14623, USA} 
\author{Jeffrey A.~Steidle}
\affiliation{Microsystems Engineering, Rochester Institute of Technology, Rochester, New York 14623, USA} 
\author{Christopher C.~Tison}
\affiliation{Air Force Research Laboratory, Rome, New York 13441, USA}
\author{Gregory A.~Howland}
\affiliation{School of Physics and Astronomy, Rochester Institute of Technology, Rochester, NY 14623, USA}
\affiliation{Microsystems Engineering, Rochester Institute of Technology, Rochester, New York 14623, USA} 
\author{Stefan F.~Preble}
\affiliation{Microsystems Engineering, Rochester Institute of Technology, Rochester, New York 14623, USA} 

\begin{abstract}
Photon pair generation in silicon photonic integrated circuits relies on four wave mixing via the third order nonlinearity. Due to phase matching requirements and group velocity dispersion, this method has typically required TE polarized light. Here, we demonstrate TM polarized photon pair production in linearly uncoupled silicon resonators with more than an order of magnitude more dispersion than previous work. We achieve measured rates above 2.8 kHz and a heralded second order correlation of $g^{(2)}(0) = 0.0442 \pm 0.0042$. This method enables phase matching in dispersive media and paves the way for novel entanglement generation in silicon photonic devices.
\end{abstract}

\maketitle

Photonic integrated circuits (PICs) provide a miniature, stable, and scalable platform for developing future light-based quantum technologies, including sensors, secure communications, and information processors \cite{Wang2019}. These systems require the on-chip generation of high quality single photons or correlated photon pairs \cite{Bruch2018}. The required on-chip photon source is bright, efficient, scalable, and produces indistinguishable or heralded photons. 

There is strong motivation to realize these quantum PICs in silicon at telecom wavelengths. Such developments can leverage both the existing CMOS electronics fabrication and manufacturing processes as well as the widespread telecommunications fiber-optic infrastructure. Consequently, significant effort has been devoted towards silicon-photonic photon source development \cite{Silverstone2016,Ma2017}. Since silicon lacks a second-order optical nonlinearity, spontaneous four-wave mixing (SFWM)---a weaker third-order effect---is utilized. In order to increase source brightness, many employ resonant structures such as micro-ring \cite{samara2019} or micro-disk resonators \cite{Lu2016} to increase the effective interaction length. While these resonant sources enhance the source brightness, precise control is required to generate and extract photon-pairs from the resonator \cite{Tison2017}. Methods to reduce parasitic processes \cite{Heuck2019} and enhance extraction efficiency \cite{Heuck2018} are active areas of research. 

Furthermore, the large polarization mode dispersion (PMD) in silicon waveguides has limited resonant photon-pair generation to just the TE polarization  \cite{Harada2010}. Xiaoge Zeng and Miloš A. Popović demonstrated dispersion engineering via the use of three coupled resonators which were independently tunable \cite{Zeng2014,Zeng2015}. Recently, Menotti \emph{et al.}~proposed a four-wave mixing (FWM) scheme in a pair of resonators that are non-linearly coupled but linearly uncoupled \cite{Menotti2019}. In this system, only one set of correlated energy modes are enhanced and able to transfer between the two resonators. They went on to experimentally demonstrate a classical, seeded device using low-dispersion TE-polarized light with nonlinear mixing between frequency modes \cite{Tan2019}.

In this letter, we experimentally demonstrate a similar dual-resonator correlated photon source for highly dispersive TM-polarized light. We produce high-quality, heralded single photons at detection rates up to 2.8 kHz with $g^{(2)}(0) = 0.0442 \pm 0.0042$ near 1550 nm. Our results show resonant enhancement of photon generation in highly dispersive media is possible, paving the way for a variety of applications. These include telecom-to-visible frequency conversion and the generation of hyper-entangled photons, with entanglement between polarization, path, energy and time.

\begin{figure*}[htbp]
  \centering
  \includegraphics[width=\textwidth]{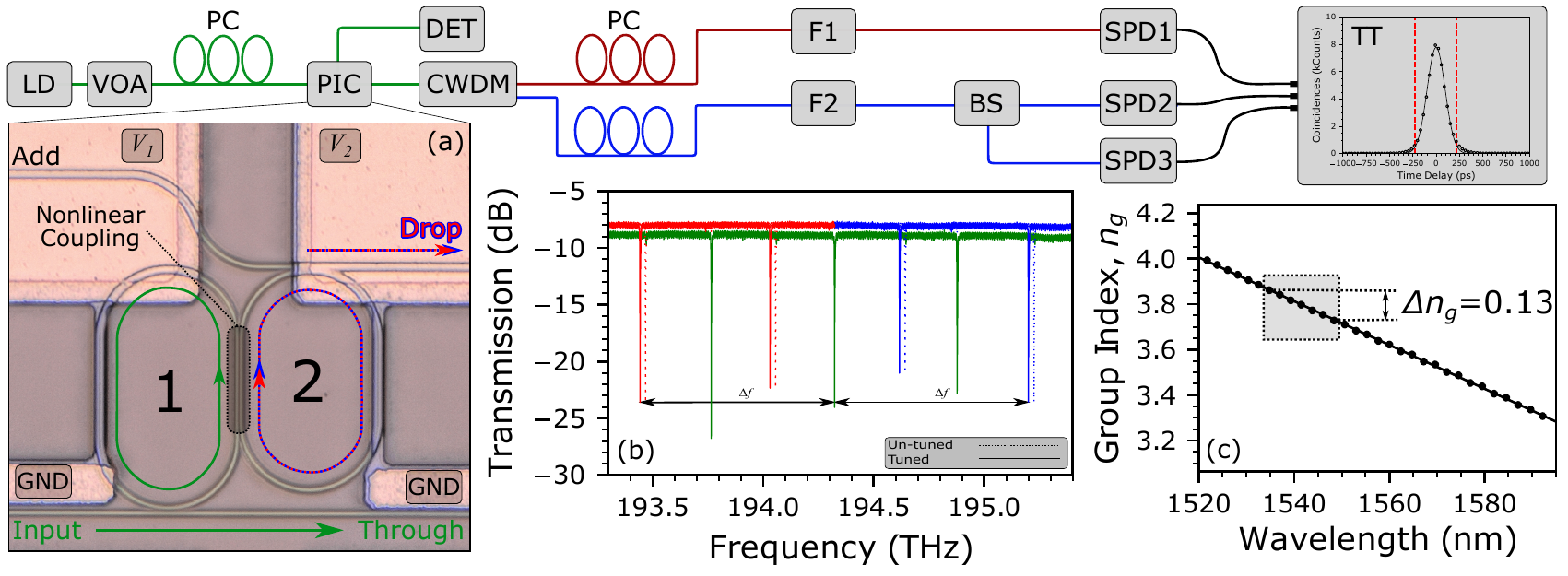}
  \caption{Experimental setup. Light from a laser (LD) passes through a variable optical attenuator, polarization controller (PC) and then enters the photonic integrated circuit (PIC). Light is collected from the PIC and sent either to a detector (DET), or through a coarse wavelength division multiplexer (CWDM) to separate signal and idler photons. Each path then passes through polarization controllers and filters (F1 and F2). Signal photons are separated at a 50:50 beam splitter (BS) and then photons are detected via superconducing nanowire  single photon detectors (SNSPDs) and correlated with a time tagger (TT). (a) PIC showing the input pump field (green) and the signal and idler photons (blue and red) created in the DC and extracted through the drop port. Voltages $V_1$ and $V_2$ are applied to tune the first and second resonators, respectively. (b) The transmission through the PIC from the input port to the through port (green) and the add port to the drop port (blue (signal side)/ red (idler side)) is shown. The green resonances correspond to resonator one and the blue/red resonances correspond to resonator two. The dashed curve shows the un-tuned resonances of resonator two, resulting in poor phase matching. The solid curve shows a symmetric resonance structure, allowing for energy conservation and efficient SFWM. (c) The averaged measured group index as determined by the free spectral range from (b). The region where the experiments are conducted is highlighted, showing the large change in group index.}
  \label{fig:exp}
\end{figure*}

We begin by considering two racetrack-style resonators that are critically coupled to separate waveguides, as shown in Fig.~\ref{fig:exp}a. Each resonator can be independently tuned via resistive heating with voltages applied at $V_1$ and $V_2$. The two resonators interact via a directional coupler (DC) which is designed to ensure input light remains within resonator one. In this way, the two resonators are linearly uncoupled. Pump light (green in the figure) was directed in through the input port, on resonance with resonator one. Due to the third order nonlinearity in silicon, signal and idler photons in a wide range of wavelengths could be generated via SFWM in the first resonator. However, in the DC, signal and idler photons that are resonant with resonator two are interferometrically enhanced, and couple into resonator two (blue/red in the figure). Therefore, the two resonators are nonlinearly coupled. We note though that SFWM will only be enhanced when the resonances are equally spaced about the pump wavelength, a strict requirement for energy conservation. However, since the two resonators can be independently controlled, it is possible to realize efficient photon-pair generation (alignment of the resonances) even in the case where the individual resonators are highly dispersive  \cite{Menotti2019, Tan2019}. 

Here we focused on non-degenerate SFWM photon-pair generation using TM polarized light, which has a large group velocity dispersion (order of magnitude larger than TE). We also obtained the advantages expected by the authors of Ref.~\onlinecite{Menotti2019}; namely, enhancement of the SFWM process, reduction of parasitic processes, compensation of self phase modulation, and less restrictive requirements on phase matching. 

We now turn to the experimental details. Light from a tunable, narrow-band laser near 1550 nm was directed through a variable optical attenuator (VOA), polarization controller and then into the PIC (see Fig.~\ref{fig:exp}). The PIC was fabricated through the Applied Nanotools Inc.~NanoSOI prototyping service and is a silicon-on-insulator (SOI) device with 500 nm $\times$ 220 nm waveguides defined through electron-beam lithography, tri-layer metalization, and oxide deposition. The 600 nm gap between the pump waveguide (input/through) and resonator one was the same as the gap between the output waveguide (add/drop) and resonator two. The directional coupler (DC), where the nonlinear interaction was enhanced, had a gap of 250 nm and was $L = 18$ $\mu$m long (designed to have zero linear coupling between the resonators). Each resonator (one and two) had a round-trip length of $\mathcal{L}_1 = 138$ $\mu$m and $\mathcal{L}_2 = 130$ $\mu$m, respectively. The PIC was secured with a temperature-controlled vacuum mount held at $27.2^\circ$C. 

Single mode fiber (SMF-28) was fusion spliced to an ultra-high numerical aperture (UHNA) fiber to improve fiber-to-PIC coupling \cite{Yin2019}, with an approximate 2.7 dB per facet fiber-to-chip loss. The through and drop ports were coupled and directed to either a high speed power meter, or through a coarse wavelength division multiplexer (CWDM), polarization controllers (PC), low-loss tunable grating filters (F1 and F2), and finally superconducting nanowire single photon detectors (SNSPDs) operating below 0.80 K. The total measured losses from the chip to the SNSPDs were approximately 9.0 dB and 5.7 dB for the signal and idler photons respectively. We performed correlation measurements using a PicoHarp 300 (for standard coincidence detection) and a Swabian Time Tagger 20 (for conditioned ${g}^{(2)}$ measurements). 

First, to characterize the resonances, we scanned the input pump wavelength and measured the transmission as shown in Fig.~\ref{fig:exp}b. Resonator one (1) and two (2) had loaded quality factors of approximately $4.1\times 10^5$ and $3.7\times 10^5$, respectively. Note that these scans were not corrected for losses, but instead measured the total round-trip loss from laser (LD) to detector (DET). Measuring the power from the input port to the through port resulted in the green curve (the pump channel), while measuring from the add port to the drop port resulted in the blue and red curves (the signal/idler channels). We saw that the two resonators were linearly uncoupled; additionally, each set of resonances had approximately -15 to -17 dB of coupling. When voltage $V_2$ was applied to the heater of the second resonator, the resonances shifted via resistive heating with minimal shift (cross-talk) on resonator one. By centering (in frequency) the resonances of resonator two around a single resonance of resonator one, we ensured energy conservation in the SFWM process. 

In what follows, we focus on photon pair generation between signal-idler pairs that are three free spectral ranges (FSR, $\Delta \lambda_{FSR}$) apart to demonstrate phase matching with large dispersion. Additionally, we note here that there is no evidence of TE polarized light found in the drop port (in neither the resonance scans nor the single photon experiments detailed below). 

Using the resonance information from Fig.~\ref{fig:exp}b, we calculated the group index dispersion as shown in Fig.~\ref{fig:exp}c using the relation $n_g = \lambda^2/\Delta \lambda_{FSR} L$. In this calculation, we have averaged the results for the two resonators due to a slight discrepancy between the two curves. This discrepancy is likely due to the varied effect of the DC on each resonator. We applied a moving average since the frequency of the peaks for each resonator are unique; however, the slope of the curve should not be affected. We saw a large change in group index with a dispersion parameter of $D = -32,000$ ps/nm-km (group velocity dispersion of -40.4 ps${}^2/$m), confirming that we analyzed and measured TM polarized light. 

Utilizing the tunability of the resonators, we generated photon pairs via the SFWM process. We measured time correlations with a bin width of 32 ps and chose a coincidence window of 448 ps, as shown in the top right of Fig.~\ref{fig:exp}. The curve shows a Gaussian fit with a FWHM of 227 ps. We optimized the photon pair count rate by varying the pump wavelength and heater temperature to account for phase matching and self phase modulation. Coincidences were calculated by summing coincidence counts within the coincidence window and subtracting accidentals from an equivalent, far away integration window. The coincidence-to-accidental ratio (CAR) was found by dividing the same two numbers. We obtained a maximum measured count rate of 2855 Hz with a CAR of 237. We calculated our Klyshko efficiency via the standard method ($\eta_K = N_{si}/N_s$) \cite{Lu2016} to reach up to 7\% when the heater is optimized. 

The pair generation rate in this system, compared to a standard microring resonator of length $\mathcal{L}_1$, should be reduced by a factor of $L/4\mathcal{L}_1 = 3.26 \%$. This is due to the shorter effective interaction length $L$ and opposite phase oscillations of the pump and signal fields \cite{Menotti2019}. Despite this, we estimated a maximum generation rate of approximately $1.3\times 10^5$ Hz by accounting for losses between the PIC and the SNSPDs. This implies that a single resonator of the same length $\mathcal{L}_1$ would result in a pair generation rate of $4.0\times 10^6$, in rough agreement with previous results \cite{Savanier2016}.

\begin{figure}[htbp]
  \centering
  \includegraphics[width=3.5in]{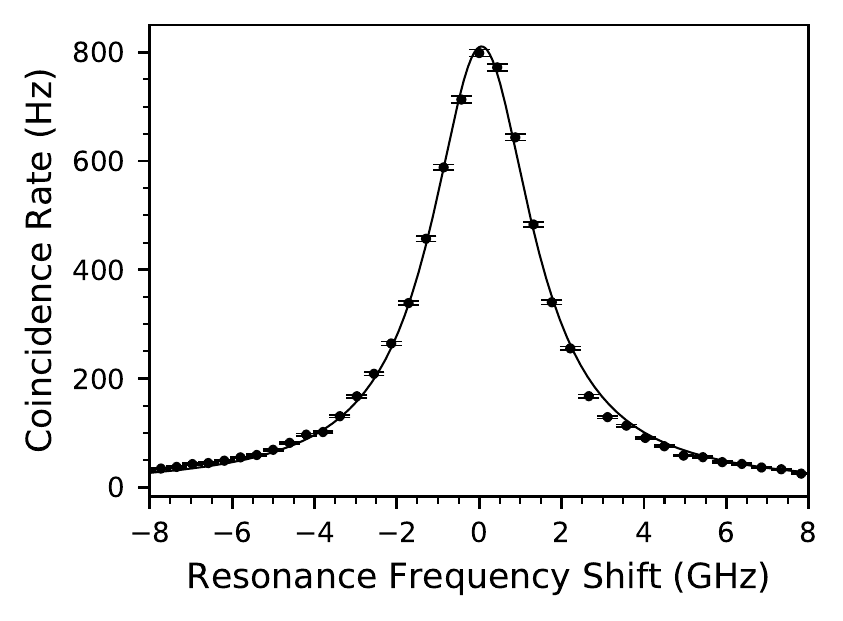}
  \caption{The location of the resonances for resonator two are scanned via resistive heating. We include a Lorentzian least squares fit with a FWHM of 2.98 GHz. Each data point is collected with 3.0 minutes of integration.}
  \label{fig:heatershift}
\end{figure}

To confirm that the dominant pair generation processes rely on the nonlinear coupling of the two resonators, we varied the heater current of resonator two in order to shift its resonances. We measured coincidences for 3 minutes at each setting and the results are shown in Fig.~\ref{fig:heatershift} with accidentals subtracted. The horizontal axis is calibrated by measuring the shift in resonances shown in Fig.~\ref{fig:exp}b as a function of current. The full width at half maximum (FWHM) of a Lorentzian least squares fit was found to be 2.98 GHz, confirming that efficient photon-pair generation requires the resonances to be precisely aligned for energy conservation. 

\begin{figure}[htbp]
  \centering
  \includegraphics[width=3.5in]{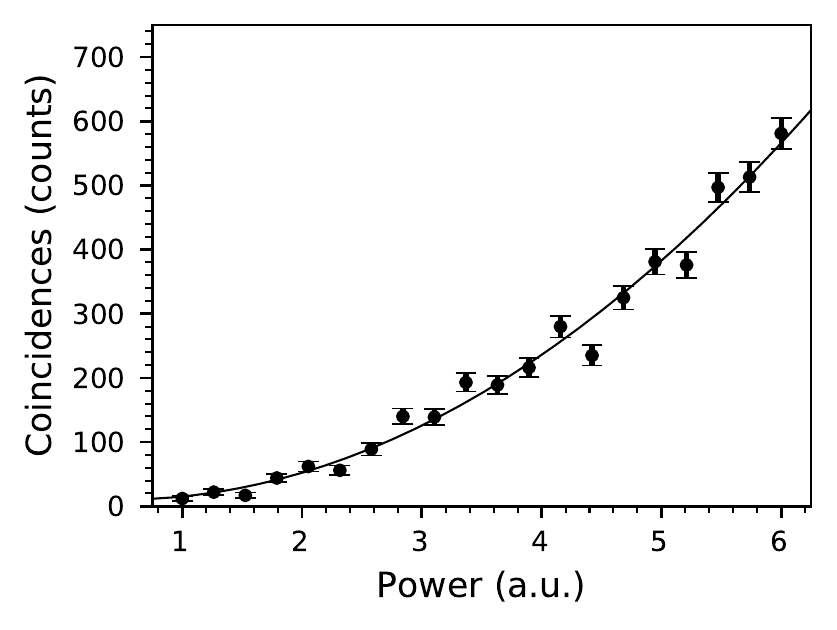}
  \caption{The input pump power was varied and coincidences were measured, exhibiting the quadratic behavior of the device. Error bars enclose one standard deviation, and the maximum input waveguide power was less than 0.5 mW, based upon loss measurements.}
  \label{fig:pumppower}
\end{figure}

Since the SFWM process is nonlinear, we varied the input power to demonstrate the expected quadratic dependence as shown in Fig.~\ref{fig:pumppower}. For each setting, the input power was set (with a maximum power of approximately 0.5 mW) and the pump frequency and resonances of resonator two were scanned to optimize coincidence counts. Then, a 20 s integration at the optimum scanned setting was acquired and accidentals were subtracted.  

\begin{figure}[htbp]
  \centering
  \includegraphics[width=3.5in]{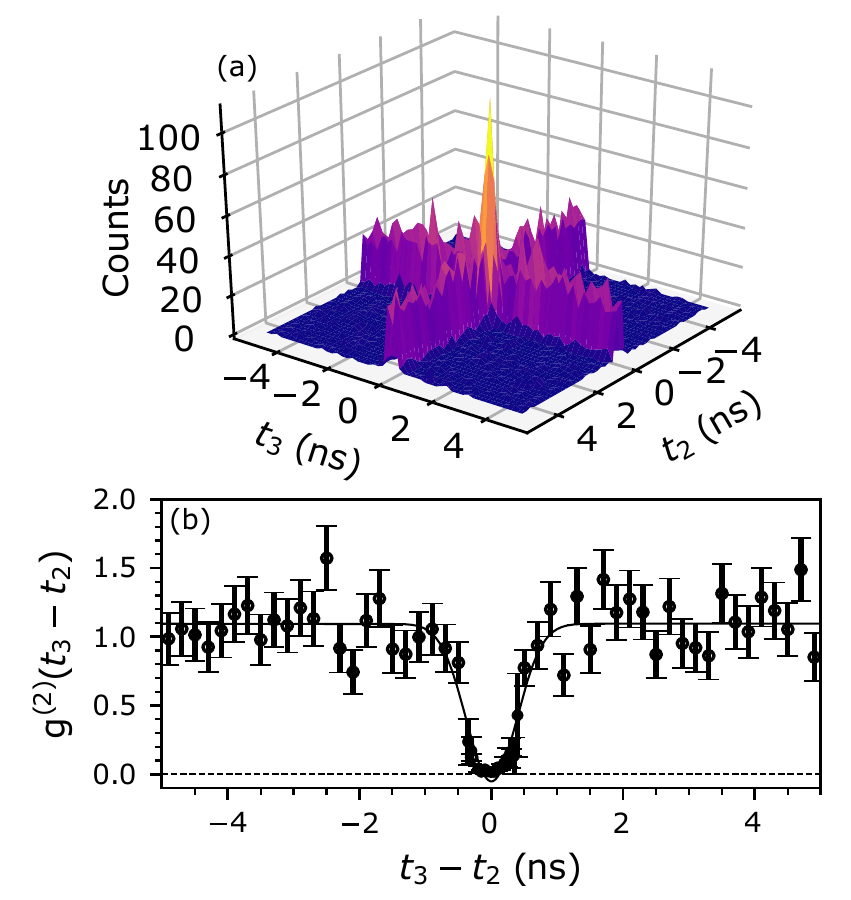}
  \caption{(a) A histogram of the number of triple coincidences between channels 1, 2 and 3, with $t_1 = 0$. (b) The value of $g^{(2)}$ as a function of the time delay between channels 2 and 3, where $t_1 = t_2 = 0$. This data includes two sets of 10 hours of integration (with 200 ps and 50 ps bin sizes) and an average coincidence rate of 437 Hz. We show only the data for the smaller bin size near $t_3 - t_2 = 0$ where triple coincidences are more common and error bars are small. A Gaussian fit to both sets of data combined gives a FWHM of 904 ps.}
  \label{fig:g2}
\end{figure}

Lastly, to analyze the quantum properties of the photon pairs, we performed a conditional second order correlation measurement ($g^{(2)}$) as shown in Fig.~\ref{fig:g2}. We accomplished this by splitting the signal photons with a 50:50 fiber optic coupler into channels 2 and 3 and measured a standard second order coherence measurement \emph{conditioned} on a detection in channel 1 \cite{Beck2007}, 
\begin{equation}
    g^{(2)}(t_3) = \frac{N_{123}(t_3) N_1}{N_{12}N_{13}(t_3)}.    
\end{equation}
We report the results by varying the time delay between channels 2 and 3 and measuring triple coincidences as shown in Fig.~\ref{fig:g2}(a) \cite{Razavi2009}. In Fig.~\ref{fig:g2}(b), we held $t_1 = t_2 = 0$ fixed and vary $t_3$; the results for varying $t_2$ instead are similar.  With approximately 2.1 mW of power in the input waveguide yielding 242 Hz coincidences, we measured a $g^{(2)}(0) = 0.0442 \pm 0.0042$, over 200 standard deviations below the classical threshold. 

In conclusion, despite large group velocity dispersion we demonstrated phase matching in the SFWM process, made possible by the relatively short interaction length in the nonlinear coupling region and the interferometric enhancement of the second resonator. The process was sensitive to the tuning of the "read-out" resonator and coincidences showed quadratic dependence on input pump power. We obtained a maximum measured count rate of 2855 Hz with a CAR of 237 using 2.1 mW of pump power in the waveguide. We obtained these results using only a single pump-rejection filter on each output channel. Furthermore, we did not employ polarization-maintaining fiber, or filter the pump to limit amplified spontaneous emission or Raman scattering. 

We have also shown single photon quantum behavior in a silicon resonator for the first time using highly dispersive TM polarized light with a group velocity dispersion more than one order of magnitude larger than previous systems (-40.4 ps${}^2/$m vs +1.33 ps${}^2/$m) \cite{Savanier2016}. We note here that the phase mismatch $\Delta$ is linear in terms of the dispersion parameter $D$ and quadratic in the detuning $\Omega$: $\Delta \approx -D \lambda^2 \Omega^2/2\pi c$ \cite{Agrawal2010}. Given that $D$ is a factor of 30 larger for TM than TE, we predict that phase matching is possible over an even larger frequency range ($\sqrt{30} \approx 5.5$, ) utilizing TE polarized light with this method. These results allow for the creation of hyper-entangled photons using polarization, path, time and energy in a silicon platform. Additionally, this process may be applicable to telecom-to-visible spectral transduction of entangled photons given the flexibility in phase matching. Lastly, we believe this work has implications for phase matching in nonlinear processes more generally.

\section*{Acknowledgements}
This work was supported by the Pennsylvania State University and the Air Force Research Laboratory (FA8750-16-2-0140). Any opinions, findings, and conclusions or recommendations expressed in this material are those of the author(s) and do not necessarily reflect the views of the AFRL.

DISTRIBUTION A.  Approved for public release: distribution unlimited. (88ABW-2019-5722)

\bibliographystyle{apsrev4-1}
\bibliography{starling2019}

\end{document}